\documentclass[journal]{IEEEtran}

\ifCLASSINFOpdf
\else
   \usepackage[dvips]{graphicx}
\fi
\usepackage{cite}
\usepackage{url}
\usepackage{amsmath,graphicx,bm}
\usepackage{color}
\usepackage{amssymb}
\usepackage{booktabs}
\usepackage{hyperref}

\usepackage{graphicx}

\newcommand{\lwlrap}{\emph{l$\omega$lrap}}

\begin{document}

\title{Addressing Missing Labels in Large-Scale Sound \\ Event Recognition Using a Teacher-Student \\ Framework With Loss Masking}

\author{Eduardo Fonseca, \IEEEmembership{Student Member, IEEE}, Shawn Hershey, Manoj Plakal, \\ Daniel P. W. Ellis, \IEEEmembership{Fellow, IEEE}, Aren Jansen, \IEEEmembership{Member, IEEE}, and R. Channing Moore
\thanks{E. Fonseca is with the Music Technology Group at Universitat Pompeu Fabra, Barcelona, Spain (eduardo.fonseca@upf.edu). This work was conducted while E. Fonseca was interning at Google Research.}
\thanks{S. Hershey, M. Plakal, D. P. W. Ellis, A. Jansen and R. C. Moore are with Google Research, New York, USA (\{shershey,plakal,dpwe,arenjansen,channingmoore\}@google.com).}}

\markboth{IEEE Signal Processing Letters, Vol. 27, 2020}
{Shell \MakeLowercase{\textit{et al.}}: Bare Demo of IEEEtran.cls for IEEE Journals}
\maketitle

\begin{abstract}
The study of label noise in sound event recognition has recently gained attention with the advent of larger and noisier datasets.
This work addresses the problem of {\em missing labels}, one of the big weaknesses of large audio datasets, and one of the most conspicuous issues for AudioSet.
We propose a simple and model-agnostic method based on a teacher-student framework with loss masking to first identify the most critical missing label candidates, and then ignore their contribution during the learning process.
We find that a simple optimisation of the training label set improves recognition performance without additional computation.
We discover that most of the improvement comes from ignoring a critical tiny portion of the missing labels.
We also show that the damage done by missing labels is larger as the training set gets smaller, yet it can still be observed even when training with massive amounts of audio.
We believe these insights can generalize to other large-scale datasets.
\end{abstract}
\begin{IEEEkeywords}
Sound event recognition, label noise, missing labels, teacher-student, loss masking
\end{IEEEkeywords}

\IEEEpeerreviewmaketitle
\vspace{-2mm}
\section{Introduction}
\label{sec:intro}
\IEEEPARstart{A}{s} SOUND Event Recognition (SER) has gained attention in recent years \cite{virtanen2018computational}, research in this field has moved from small datasets encompassing few hours of audio \cite{foster2015chime,piczak2015esc,salamon2014dataset,fonseca2018general}, to larger datasets with much greater coverage and duration \cite{gemmeke2017audio,Fonseca2019audio}.
A milestone was the release of AudioSet---a dataset of 527 everyday sound classes organized with a hierarchical ontology, that includes around 2 million segments of $\approx$10s in its released version \cite{gemmeke2017audio}.
However, large-scale audio datasets inevitably bring in label noise issues, since it is intractable to exhaustively annotate such massive amounts of audio. The resulting issues of less-precise labels can cause various problems including performance decreases and longer training times \cite{frenay2014classification}, and can become a critical impediment to the success of SER.
Consequently, label noise in SER has lately become a focus of interest. 
Previous work analyses the impact of label noise in these tasks \cite{meire2019impact,shah2018closer}, as well as proposes ways to mitigate its negative effect \cite{Fonseca2019learning,Fonseca2019model, kumar2018learning,kumar2019secost}. A DCASE 2019 Challenge Task was launched to foster research in this topic\cite{Fonseca2019audio}.

AudioSet presents a number of label noise problems.
Some are due to shortcomings in the annotation process, e.g., missing or incorrect labels.
Others are related to the hierarchical structure of the AudioSet Ontology, e.g., a segment may be annotated with a leaf class label but not with its parent one, or annotated with a label that is not the most specific within its hierarchical path.
Still other problems arise from the temporally-weak labels (i.e., clip-level labels), where the class label may be active only during a small (and unknown) portion of the audio segment. 
Finally, some semantic inconsistencies may exist as the ontology allows for several sound attributes to be associated to one type of sound event (while not all of them may have been annotated).
Despite these label noise problems, they have been directly addressed in only a few of the previous works using AudioSet (e.g., \cite{kumar2018learning,kumar2019secost}), while the majority of efforts focus on deriving more sophisticated network architectures that ignore or downplay the idiosyncrasies of the labeled audio data (e.g., \cite{kong2019panns,ford2019deep}).

In this work, we focus on one of the most frequent label noise problems in AudioSet: its missing labels.
The study of missing labels in SER has received very little attention. 
To our knowledge, this specific topic has been covered only by \cite{meire2019impact}, where robustness to missing labels is studied by simulating them in a synthetic dataset of 20 classes.
Our contribution is two-fold.
First, we propose a simple and model architecture-agnostic method based on a teacher-student framework to first identify the most critical potentially missing labels in AudioSet, and then ignore their contribution in the learning process through a loss masking approach.
We then analyse the effect of the proposed method via a set of experiments using two model architectures of different capacity and two train sets of different size.
We find that a simple optimisation of the training label set can lead to a non-negligible improvement in recognition performance without additional compute.
We also discover that most of the improvement comes from ignoring a tiny portion of the missing labels. 
The ultimate goal is to demonstrate how prior knowledge of a dataset can be leveraged to build simple, efficient, and model-agnostic solutions to improve recognition performance, which can complement other approaches focused on improving network architectures.


\vspace{-1mm}
\section{Missing Labels in AudioSet}
\label{sec:missing}
\begin{figure*}[ht]
    \vspace{-4mm}
    \centering
  \centerline{\includegraphics[width=1.02\textwidth]{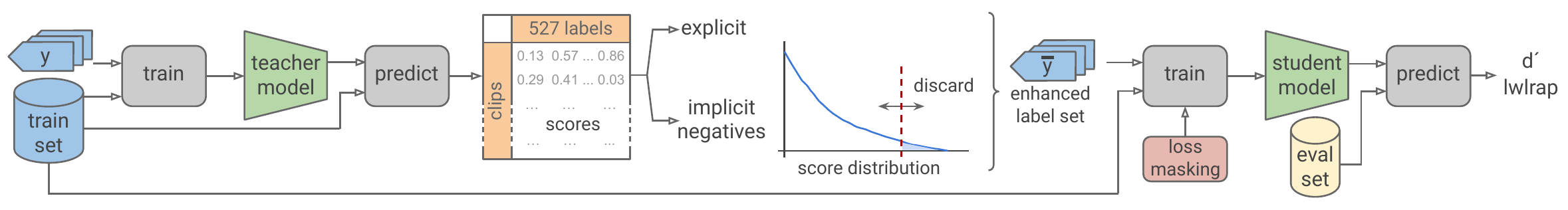}}
     \vspace{-3mm}
    \caption{Proposed method. First: Identification of potential missing labels per class using teacher's predictions and creation of enhanced label set. Second: Training a student model while ignoring missing labels through loss masking.}
    \label{fig:method}
    \vspace{-4mm}
\end{figure*}
We refer to missing labels as those labels that would be included in an ideal, exhaustive annotation but which are missing from the current set.
The existence of missing labels in AudioSet is due to the dataset curation process.
This process consisted of two steps: the compilation of a list of candidate labels per clip, and the human validation of the labels nominated in that list.
The list of candidate labels was compiled by means of a series of automatic methods, including the processing of the available metadata (e.g., video title and/or description) as well as a query-by-example method.
These methods can be sub-optimal due to the high inter- and intra-class variation of sound events in the AudioSet Ontology \cite{gemmeke2017audio}. 
In addition, the list of candidate labels was limited to a maximum of ten labels per clip.
There are therefore several ways by which some existing sound events fail to be nominated by the system, or are nominated but ranked below the top ten, thus leading to missing labels.
We call the nominated labels that have received human validation \textit{explicit} labels (that can in turn be positive or negative, depending on the human rating being ``Present" or ``Not Present").
The remaining labels which are not proposed by the nomination system (the vast majority) are referred to as \textit{implicit} negative labels, and have received no human validation.
Hence, it is likely that some of the implicit negative labels are indeed missing (positive) labels.

AudioSet poses a multi-label audio tagging problem, which is usually addressed by a deep network with an output layer composed by $C$ binary classifiers, with $C$ being the number of classes.
Binary classification loss functions are typically adopted, composed by two terms, one accounting for the positive examples, and the other for the negative ones. The default option is binary cross-entropy, expressed by: \eqref{eqn:bce}
\vspace{-1mm}
\begin{equation}
\label{eqn:bce}
\mathcal L=-\sum_{c=1}^C y_c\log(p_c) + (1-y_c)\log(1-p_c),
\vspace{-1mm}
\end{equation}
where $p_c$ represents the network output prediction and $y_c$ the ground truth label for class $c$.
The implicit negative labels are considered negative examples, hence they are covered by the second term of \eqref{eqn:bce}.
If a sound event is actually present, we want the model to emit a high score even if the ``Present'' label is missing.
However, this behaviour will be penalized (with the penalty increasing for higher output predictions) due to the backpropagation of an artificially high loss contribution, which causes a misleading gradient update. 
We hypothesize this hinders the learning process to some extent.

\vspace{-2mm}
\section{Method}
\label{sec:method}

We propose a two-step strategy based on a teacher-student framework \cite{ba2014deep} depicted in Fig.~\ref{fig:method}.
First, a teacher model is trained using the original AudioSet labels, $\bm{y}$, and the teacher’s predictions are used to build a new enhanced label set, $\bm{\bar{y}}$, where the suspected missing labels are flagged.
Second, $\bm{\bar{y}}$ is used to train a student model where the flagged labels are ignored through a loss masking approach.
Next we explain the proposed method in detail.
The first step consists of identifying the potential missing labels per class.
To do so, a \textit{teacher} model is trained using the original AudioSet labels, $\bm{y}$.
We use the trained teacher model to predict scores for the train set, leading to a set of $\mathbb{R}^{C\times1}$ scores per audio clip.
The teacher's predicted scores are used to take decisions on labels' veracity.
We focus on the predictions associated with the implicit negative labels.
Our hypothesis is that the top-scored implicit negative labels (henceforth, \textit{top-scored negatives}) are likely to correspond mostly to missing ``Present'' labels, i.e., false negatives.
Under this hypothesis, we rank implicit negative labels based on the teacher's predictions and we create a new label set, $\bm{\bar{y}}$, by flagging a given percentage of the top-scored negatives per class, with the intention of ignoring them in the student’s learning.
Note that, unlike other teacher-student pipelines where teacher’s predictions are used as ground truth to train a student (e.g., via soft labels \cite{ba2014deep,li2014learning}), our case features a \textit{skeptical} teacher whose supervision is used to highlight flaws in the current ground truth, estimating potentially missing labels and flagging them in a new label set. 
The outcome of this step is an enhanced training label set, $\bm{\bar{y}}$, where the label information is encoded as multi-hot target vectors with three states (positive, negative, and to-be-ignored labels).

The second step consists of training a \textit{student} model using the label set optimised through the teacher’s predictions. 
The goal here is to ignore the loss contributions of the previously flagged labels in the loss function computation.
This is done through a loss masking approach, where we modify the student's learning pipeline so as to create a binary mask of size $C\times1$ per input example, using the information encoded in $\bm{\bar{y}}$.
Each element of the binary mask, $M_c$, is defined as \eqref{eqn:mask}
\begin{equation}
\label{eqn:mask}
M_c=\begin{cases}
    0, & \text{if label is implicit negative and score $>t_c$ }\\
    1, & \text{otherwise},
\end{cases}
\vspace{-1mm}
\end{equation}
where $t_c$ is a per-class threshold computed as a given percentile of the per-class scores distribution.
In practice, we compute the loss function $\mathcal L$ following \eqref{eqn:bce}, and $M_c$ is applied to the negative term of $\mathcal L$ in order to discard the loss contributions of potentially missing labels.
A similar masking approach to ignore false negatives in SER was recently used in~\cite{kim2019sound}.
\vspace{-1mm}
\section{Experimental Setup}
\label{sec:setup}
We evaluate the proposed method using an internal version of AudioSet \cite{gemmeke2017audio}, including information about which labels are explicit/implicit.
To study the impact of missing labels as a function of training data size, we use two train sets with similar class distributions but one roughly five times larger than the other (see Table~\ref{tab:train_and_models}).
\begin{table}[!t]
\vspace{-2mm}
\caption{Train sets and architectures used in our experiments}
\vspace{-2mm}
\centering
\begin{tabular}{lcc}
\toprule
\textbf{Train set} & \textbf{clips} & \textbf{hours}\\
\midrule
\textit{tr\_small}       & 506,721	        & 1407     \\
\textit{tr\_large}       & 2,467,357	    & 6853       \\
\midrule
\textbf{Architecture} & \textbf{parameters} & \textbf{Mult-Adds}\\
\midrule
ResNet-50          & 30M	        & 1860M     \\
MobileNetV1       & 3.7M	    & 69.2M      \\
\bottomrule
\end{tabular}
\label{tab:train_and_models}
\vspace{-4mm}
\end{table}
For evaluation, we use an internal \textit{eval set} of 47,132 audio clips.
Incoming clips are transformed to log-mel spectrograms using a 25ms Hann window with 10ms hop, and 64 mel log-energy bands.
The network is presented with time-frequency patches of 96 frames (corresponding to 0.96s) with 50\% overlap.
To assess the impact of missing labels on models of different capacity, we employ two 
CNNs as students: ResNet-50 \cite{he2016deep} and MobileNetV1 \cite{howard2017mobilenets}.
Both are taken from the computer vision literature and have proven successful in audio recognition (see \cite{hershey2017cnn} and YAMNet\footnote{\url{https://github.com/tensorflow/models/tree/master/research/audioset/yamnet}\label{footnote_YAMNet}}).
Table~\ref{tab:train_and_models} shows model size and Mult-Adds for both architectures.
We use Adam optimizer \cite{kingma2014adam} with learning rate 1e-5, 
and random weight initialization with a standard deviation of~0.001.

\vspace{-2mm}
\subsection{Evaluation}
\label{ssec:eval}
We pass each 0.96s evaluation patch through the model to compute output scores, which are then averaged per-class across all patches in a clip to obtain clip-level predictions, as in \cite{gemmeke2017audio}.
As evaluation metrics we use $d’$ and {\lwlrap}. 
$d’$ (d-prime) is a within-class metric, i.e., it ranks all test samples according to the classifier score for a given class. 
$d’$ can be computed as a monotonic transform of ROC AUC, and describes the separation between unit-variance normal distributions that would achieve the same AUC \cite{hershey2017cnn,green1966signal}.
In order to avoid the impact of potential missing positive labels in the eval set, only samples with explicit labels for a given class (both positive and negative) are used in the calculation of $d'$. (Because this excludes many ``easy'' samples, the resulting $d'$ values are substantially lower than including all non-positive samples as negatives.) 
Label-weighted label-ranking average precision ({\lwlrap}) is a between-class metric, i.e., it evaluates the overall ranking across all classifier outputs for every test sample.
Specifically, {\lwlrap} measures, for every ground truth test label $c$, what fraction of the predicted top-ranked labels down to $c$ are among the ground truth\cite{Fonseca2019audio}.
Both metrics are computed on a per-class basis, then averaged with equal weight across all classes to yield the overall performance shown in Table~\ref{tab:results} and Fig.~\ref{fig:perf_vs_discard}. Non-pathological
$d’ \in [0, \infty)$ while {\lwlrap}  $\in [0, 1]$.

\section{Experiments}
\label{sec:experiments}
\begin{figure}[t]
  \vspace{-4mm}
  \centering
  \centerline{\includegraphics[width=1.03\columnwidth]{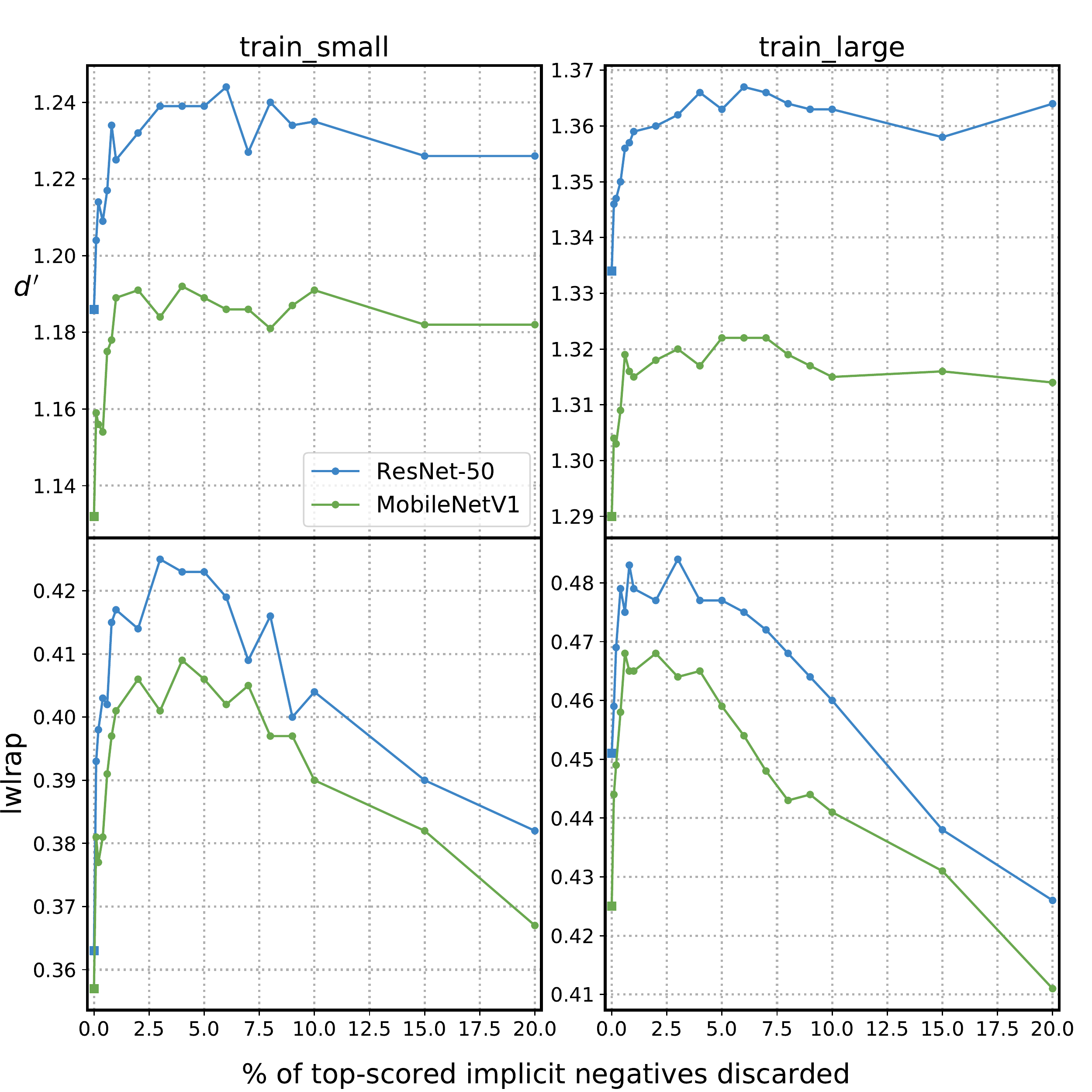}}
  \vspace{-1mm}
  \caption{Classification performance as a function of the proportion of  top-scored negative labels that are discarded. Each point in the lines corresponds to one operating point. The leftmost point in each curve, 
  marked with a square, corresponds to using all negative labels.}
  \label{fig:perf_vs_discard}
  \vspace{-5mm}
\end{figure}
As explained in Section \ref{sec:method}, we first train a teacher model with the unmodified labels and use it to predict scores in the train set. 
We used an internal ResNet-50 model for the teacher which had been trained using several tweaks to improve performance, similar to those used in the publicly-released YAMNet model.\textsuperscript{\ref{footnote_YAMNet}} 
Based on the teacher’s predicted scores, we generate 18 new label sets, each of them using a different threshold $t_c$, i.e., discarding a different proportion of top-scored negatives in the train set.
Finally, for every enhanced label set, we train a student model on the train set, and predict on the evaluation set, reporting the best performance obtained.\footnote{While choosing parameters based on the test set introduces overfitting, our experience with data at this scale (i.e., validation and test sets in the range of hundreds of hours) is that results obtained by this suspect methodology are in practice similar to those from a more rigorous separation of tuning and evaluation sets.}

Fig.~\ref{fig:perf_vs_discard} illustrates the impact of missing labels by plotting performance 
as a function of the amount of top-scored negatives discarded, similar to the treatment of noisy ImageNet labels in \cite{northcutt2019confident}.
We experiment with progressively discarding $t_c \in \left\lbrace 0, 0.1, 0.2, 0.4, 0.6, 0.8, 1, 2, 3, 4, 5, 6, 7, 8, 9, 10, 15, 20\right\rbrace$~\% of top-scored negatives for the two train sets and architectures mentioned.
Each point in the lines is the result of one experiment trial using one label set with a given amount of discarded negatives. 
The leftmost point (at $x=0.0$\%, marked with a square) corresponds to \textit{normal training} (no labels ignored and all false negative labels included).
We use it as our baseline.
Common to all the curves of Fig.~\ref{fig:perf_vs_discard}, we observe three regions from left to right: a steep increase at the beginning of the curve, followed by a sweet-spot, and a final decay that is more severe in {\lwlrap} than in \textit{d'}.
A possible interpretation of this behaviour is as follows.
We conjecture that the top-scored negatives correspond either to missing ``Present'' labels (i.e., false negatives (FNs)), {\em or} they are ``decoys'', difficult (and thus informative) true negatives (TNs), perhaps from similar classes, and especially useful in learning.
First, we remove some critical FNs that damage the learning process, hence the sudden performance increase at the left of the curves.
As we continue discarding more top-scored negatives, we keep removing FNs, but we also start to remove some TNs. 
Therefore, performance increases more slowly, until a sweet-spot is reached where both effects cancel out.
Finally, if we keep ignoring more top-scored negatives, performance is degraded.
As to why the decay in \textit{d'} is much less pronounced than in {\lwlrap}, a possible explanation lies in the way \textit{d'} works.
\textit{d'} characterizes the separation of the positive and negative score distributions as the distance between their \textit{means}.
It may be that removing the high scoring tail changes the mean of the negative distribution (hence \textit{d'} increases suddenly), but as we remove more labels with much more frequent scores the mean of the negative distribution barely changes (and consequently so \textit{d'}).
By contrast, {\lwlrap} does not suffer from this issue, its curves showing a decay as expected. 
Based on this intuition, we consider the right end of the \textit{d'} curves less reliable.
Table~\ref{tab:results} lists the performances for baselines and best operating points for all the train sets and architectures considered. 
Based on 
Fig.~\ref{fig:perf_vs_discard} and Table~\ref{tab:results}, next we make a number of observations.
\begin{table}[!t]
\vspace{-2mm}
\caption{Classification performance for baselines and best operating points for architectures and train sets considered}
\vspace{-1mm}
\centering
\begin{tabular}{@{}lccccc@{}}
\toprule
\textbf{Model} & \textbf{Train set} & \multicolumn{2}{c}{\textbf{\textit{d'}}}   & \multicolumn{2}{c}{\textbf{{\lwlrap}}} \\
                    &                    	    & baseline & best       & baseline & best \\
\midrule
ResNet-50           & \textit{tr\_small}	    & 1.186 & 1.244       & 0.363 & 0.425 \\
                    & \textit{tr\_large}	    & 1.334 & 1.367       & 0.451 & 0.484 \\
\midrule
MobileNetV1         & \textit{tr\_small}	    & 1.132 & 1.192         & 0.357 & 0.409 \\
                    & \textit{tr\_large}	    & 1.290 & 1.322         & 0.425 & 0.468 \\
\bottomrule
\end{tabular}
\label{tab:results}
\vspace{-5mm}
\end{table}

\textbf{Effect of ignoring the highest ranked top-scored negatives.}
The proposed method yields performance improvements in all cases considered.
The best operating points are usually between 3 and 6\% discarded for \textit{d'}, and between 1 and 4\% for {\lwlrap}. 
We believe this result is relevant as AudioSet training examples are often treated as if they had complete labels.
However, the most important pattern we observe in all cases is the consistent steep increase at the beginning of the curves.
In all cases, most of the improvement comes from removing just $\approx$1\% of the top-scored negatives.
Further, in most cases, just by removing a tiny percentage ($<=$0.2\%) of (potentially) missing labels, approximately half of the total boost is already attained.
Two observations can be made from these findings.
First, this indicates that a tiny portion of labels is troublesome and it is moderately affecting classifier performance, a concept which is basis for disciplines like instance selection, where it is assumed that not all training examples are equally informative, some of them being redundant and some being harmful \cite{liu2002issues}.
Second, these findings become interesting as they contrast with the common trend of acquiring more and more training data to improve recognition performance, even if noisily labeled \cite{sun2017revisiting} (something we also find useful in our experiments in general).

\textbf{Effect of train set size.} 
Table~\ref{tab:results} shows improvements with respect to the baseline of $\approx$0.060 for \textit{d'} when training with \textit{tr\_small} for both architectures, whereas when using \textit{tr\_large}, improvements are almost half of that ($\approx$0.033). 
This relationship also holds for {\lwlrap} when using ResNet-50,\footnote{By chance, absolute improvements for both metrics are numerically similar in this case, despite the metrics and their numeric range are different.}
whereas when using MobileNetV1, the performance difference between training with \textit{tr\_small} and \textit{tr\_large} is smaller.
These results seem to indicate that the damage done by missing labels, and consequently the performance boost obtained by discarding them, can be higher when the dataset is smaller.
A possible explanation is that larger amounts of data help to mitigate the effect of these errors in the label space, which accords with \cite{sun2017revisiting}. 
However, even when training with massive amounts of audio (almost 7000h, see Table \ref{tab:train_and_models}), the impact of these labelling errors can still be observed.
The \textit{d'} sweet spot occurs roughly in the same region for both train sets.
The {\lwlrap} sweet spot seems to move slightly to minimal discards when training with larger amounts of data.

\textbf{Effect of model architectures.}
The proposed method is effective for both model architectures despite having different underlying principles and significantly different numbers of parameters.
The overall trend of the curves in Fig.~\ref{fig:perf_vs_discard} is similar for both architectures.
As can be seen in Table~\ref{tab:results}, in terms of \textit{d'}, both architectures show very similar improvements with respect to their corresponding baselines.
In terms of {\lwlrap}, however, results are inconsistent, with ResNet-50 providing a greater improvement than MobileNetV1 when training on \textit{tr\_small}, and vice versa when training on \textit{tr\_large}.
We do not observe consistently larger improvements using ResNet-50, even though its much larger number of parameters might lead one to expect it to overfit labeling errors more readily. 
Regardless of missing labels, when comparing baselines, ResNet-50 outperforms MobileNetV1, but not by a particularly large margin considering the huge difference in parameters.

\textbf{Effect on evaluation metrics.}
By looking at Table~\ref{tab:results}, it can be seen that \textit{d'} improvements reach up to relative 5.3\% (MobileNetV1) and {\lwlrap} improvements reach up to relative 17.1\% (ResNet-50), both cases occurring when using \textit{tr\_small} ($\approx$ half a million clips), where improvements are more evident.


Finally, we carried out a small informal listening test in which we inspected some of the clips associated with the discarded top-scored negatives for a few classes.
As expected, most clips were missing ``Present'' labels, some of them being flagrant labelling errors, but difficult to detect considering the train set size.
These findings indicate that the proposed method, while simple, is effective in identifying missing labels in a human annotated dataset like AudioSet, and it is able to improve training over unnoticed missing labels.
Additionally, it can be useful for dataset cleaning or labeling refinement. 
Re-labelling a small amount of flagged top-scored negatives may lead to even better results than the proposed method.
While the presented results are specific to AudioSet, the insight found can also apply to other large-scale audio datasets, especially those annotated via human validation of sub-optimally nominated candidates, e.g., the FSD dataset\cite{Fonseca2017freesound}.


\vspace{-1mm}
\section{Conclusion}
\label{sec:conclusion}
We have identified missing labels as a pathology in the labelling of AudioSet.
We have proposed a simple method based on a teacher-student framework with loss masking to first identify the most critical potentially missing labels, and then ignore them during the learning process.
Our main findings are:
\textit{i)} most of the improvement comes from filtering out a tiny portion ($<$1\%) of the most critical estimated missing labels, showing a moderate impact on performance;
\textit{ii)} the damage done by missing labels (and the performance boost obtained by discarding them) becomes higher as the train set gets smaller---however, even when training with massive amounts of audio, the impact of these labelling errors can still be observed;
\textit{iii)} when applied to two CNN architectures of different nature and size the proposed method behaves similarly in both cases.
We believe these insights will apply equally to large-scale audio datasets beyond AudioSet, since the problem of missing labels is endemic.

%
\bibliographystyle{IEEEtran}
\bibliography{refs}

\newpage
\section{Supplemental Material}
\subsection{Formulating the Method from Knowledge Distillation}
Another way to introduce the proposed method is from the perspective of knowledge distillation \cite{hinton2015distilling}.
A typical formulation of distillation is $\mathcal L(p_{teacher}, p_{student})$, with $\mathcal L$ given by \eqref{eqn:bce}, whereas our method could be formulated as $\mathcal L(f(p_{teacher}), p_{student})$.
For $f$ identity, standard distillation is recovered.
We define one instantiation of $f$ as a nonlinear transform applied to the teacher scores for the implicit negatives---our \textit{skeptical} teacher. 
Similarly, other classes of $f$ might also be relevant to accommodate label noise.

\subsection{An Example of Applying the Method}
For easier assimilation, we provide an example of application of the proposed method. Table~\ref{tab:operating_point} illustrates the details of the operating point of 0.1\% discard in \textit{tr\_small} for the \emph{Ambulance (siren)} and \emph{Speech} classes. 
The total number of labels is the number of clips in the train set (506,721). 
The number of explicit labels (i.e., human rated, which are both positive and negative) is usually a tiny portion, in the range of a few hundreds or thousands (as in \emph{Ambulance (siren)}), except for a few high prior classes such as \emph{Speech}.
Implicit labels (all negative) form the remainder of the clips.
Note that \emph{Ambulance (siren)} represents the typical case that holds for the vast majority of classes, while \emph{Speech} represents an extreme case relevant to only a handful of classes.
In this operating point, we ignore the top-scored 0.1\% of the implicit negatives, which is usually around 500 labels per class, except for the few high prior classes, in which it is less.
\begin{table}[ht]
\vspace{-1mm}
\caption{Label counts for two example classes at one operating point of Fig.~\ref{fig:perf_vs_discard} (\textit{tr\_small} and discarding 0.1\% of top-scored negatives)}
\centering
\begin{tabular}{l|ccc|c@{}}
\toprule
\textbf{Class}  & \textbf{Total} & \textbf{Explicit} & \textbf{Implicit} & \textbf{To ignore} \\
\midrule
Ambulance (siren)  & 506,721              & 1657              &505,064           & 504    \\
Speech             & 506,721              & 464,262           &42,459            & 42    \\
\bottomrule
\end{tabular}
\label{tab:operating_point}
\vspace{-1mm}
\end{table}

\subsection{Per-class l$\omega$lrap analysis}
\label{ssec:scatter}
\begin{figure}[]
  \vspace{-5mm}
  \centering
  \centerline{\includegraphics[width=\columnwidth]{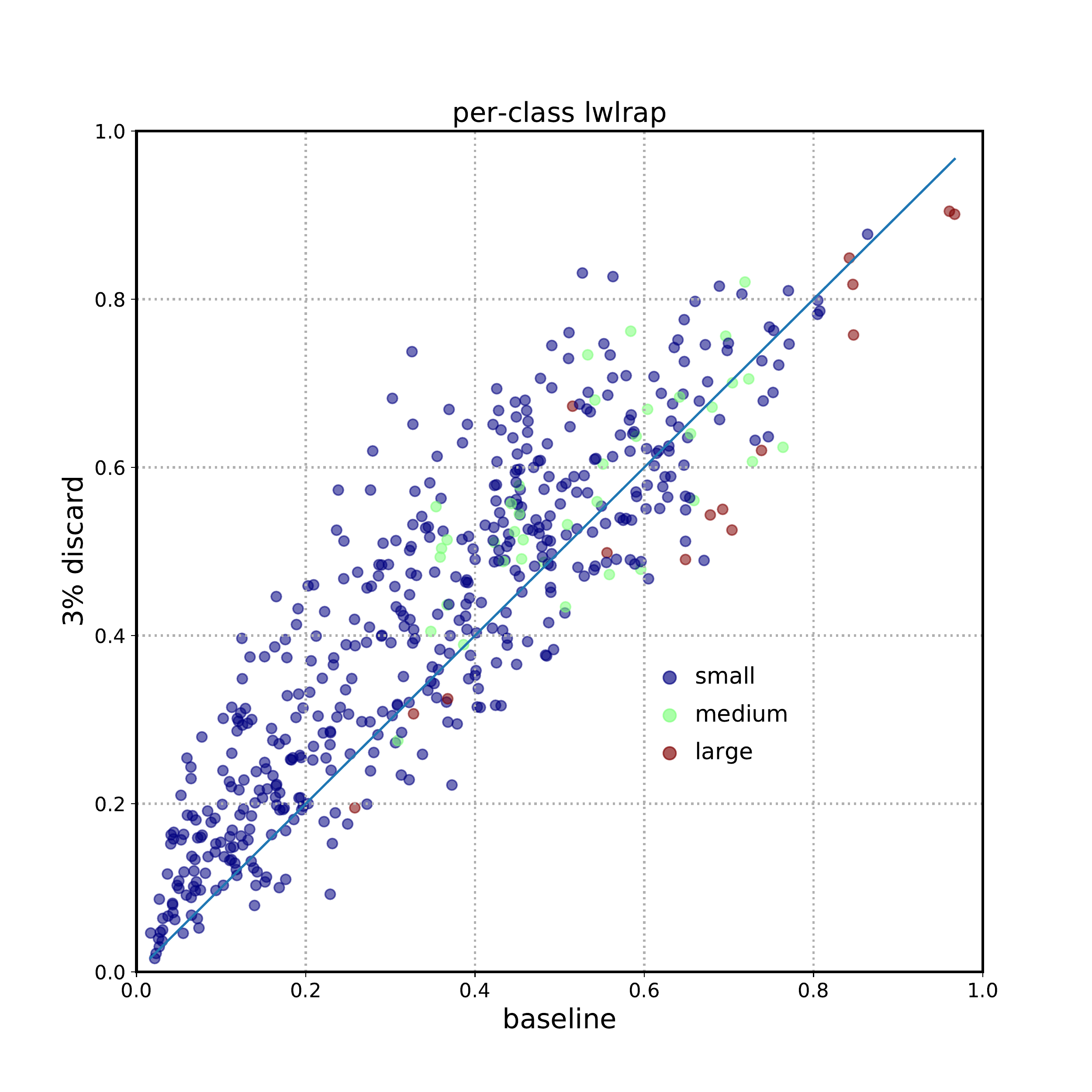}}
  \vspace{-5mm}
  \caption{Per-class {\lwlrap} for baseline (no label rejection) vs. best operating point (3\% discard) for ResNet-50 on \textit{tr\_small}.}
  \label{fig:scatter}
  \vspace{-3mm}
\end{figure}

We provide a brief per-class analysis to see how the proposed method affects the classes as a function of their prior in the dataset.
As a case study we focus on {\lwlrap} since the improvements are observed more easily, and we compare the baseline with the best operating point of ResNet-50 on \textit{tr\_small}.
Fig.~\ref{fig:scatter} shows the scatter plot of per-class {\lwlrap} values for baseline (i.e., no label rejection) vs. those of the best operating point (3\% discard); the diagonal line divides the space into classes improved (above the line) or worsened (below the line) by discarding.
We divide the 527 AudioSet classes into three groups according to their prior: \textit{i)} 15 largest classes with prior $\rho_c > 0.01$ (red), \textit{ii)} 474 smallest classes with a prior $\rho_c < 0.00325$, corresponding, approximately, to the subset of 474 leaf nodes in the hierarchy of the AudioSet Ontology \cite{gemmeke2017audio} (blue), and \textit{iii)} remaining 38 classes of medium size (green).
\newpage
Table~\ref{tab:per_class} lists the number of classes in which performance improves, along with the average improvement, for every group of classes.
\begin{table}[ht]
\vspace{-2mm}
\caption{Number of classes with improvement and average improvement for the three groups of classes in Fig.~\ref{fig:scatter}}
\centering
\begin{tabular}{l|ccc}
\toprule
\textbf{Group}  & \textbf{classes} & \textbf{classes w/} & \textbf{Avg {\lwlrap}}  \\
                 &                  & \textbf{improvement} & \textbf{improvement}  \\
\midrule
large             & 15           &2  (13.3\%)          & 0.082    \\
medium            & 38           &27 (71.1\%)           & 0.086    \\
small             & 474          &359 (75.7\%)           & 0.106    \\
\bottomrule
\end{tabular}
\label{tab:per_class}
\end{table}
In light of Fig.~\ref{fig:scatter} and Table~\ref{tab:per_class}, we see the following.
Classes with high prior tend to get slightly worse.
While the performance changes observed are relatively small, this is is somewhat surprising as the number of labels ignored is even smaller in these cases---a possible explanation is that most of the labels being discarded correspond to informative TNs.
On the contrary, groups of classes with medium and small priors present a similar percentage of classes showing improvement, being slightly larger in the group of small classes.
In addition, the average improvement is also higher in the group of small classes, with an absolute difference of 0.02.
While a more in-depth study is needed before making stronger claims, results seem to indicate that the impact of missing labels (and of the proposed method) is greater on classes with low prior, which goes in line with findings in Section \ref{sec:experiments} about the effect of train set size.
Classes that benefit the most out of this process are: \emph{Waterfall}, \emph{Fusillade}, \emph{Sizzle} and \emph{Babbling}, featuring improvements greater than 0.3.
The procedure carried out can be useful to detect classes with labelling errors, applicable in dataset cleaning or labeling refinement.

\end{document}